\documentclass[%
reprint,
 amsmath,amssymb,
aps,
pra,
]{revtex4-1}

\usepackage{graphicx}
\usepackage{dcolumn}
\usepackage{bm}
\usepackage{epstopdf}

\begin{document}


\title{Measurement of the scalar polarizability of the indium $6p_{1/2}$ state using two-step atomic-beam spectroscopy}
\author{Benjamin L. Augenbraun} \altaffiliation{Current address: Dept. of Physics, Harvard University, Cambridge, MA 02138}
\author{Allison Carter} \altaffiliation{Current address: Dept. of Physics, Univ. of Maryland, and Joint Quantum Institute, College Park, MD, 20740}
\author{P. M. Rupasinghe}
\author{P. K. Majumder}
\email{pmajumde@williams.edu}
\affiliation{Department of Physics, Williams College, Williamstown, MA 01267}
\date{\today}%


\begin{abstract}

\noindent We have completed a measurement of the Stark shift within the $^{115}$In $6s_{1/2} \rightarrow 6p_{1/2}$ excited-state transition using two-step laser spectroscopy in an indium atomic beam. Combining this measurement with recent experimental results we determine the scalar polarizability, $\alpha_{0}$, of the $6p_{1/2}$ state to be $7683 \pm43 \,a_{0}^{3}$ in atomic units, a result which agrees very well with recent theoretical calculations. In this experiment, one laser, stabilized to the $5p_{1/2} \rightarrow 6s_{1/2}$ 410~nm transition, was directed transversely to the atomic beam, while a second, overlapping laser was scanned across the 1343~nm $6s_{1/2} \rightarrow 6p_{1/2}$ transition. We utilized two-tone frequency-modulation spectroscopy of the infrared laser beam to measure the second-step absorption in the interaction region, where the optical depth is less than 10$^{-3}$. In the course of our experimental work we also determined the hyperfine splitting within the $6p_{1/2}$ state, improving upon the precision of an existing measurement.

\end{abstract}

\pacs{Valid PACS appear here}
\maketitle


\section{Introduction}
High-precision atomic structure measurements provide important tests of the accuracy of on-going efforts to calculate atomic wavefunctions in multielectron atomic systems. While this experiment-theory interplay has been especially productive in alkali-metal systems, recent advances in \textit{ab initio} calculation techniques have provided theoretical results of improved precision for multiple-valence-electron systems~\cite{Safronova2013, Safronova2009, Safronova2007, Safronova2006, Sahoo2011}. These calculations are particularly relevant for group IIIA systems such as indium and thallium. Both of these elements are of interest for fundamental physics: thallium has been an important test system for violations of discrete symmetry~\cite{Regan2002, Vetter1995, Porsev2012}, and indium has recently been proposed as a potential system in which to measure a permanent electric-dipole moment~\cite{Sahoo2011b}. The size of these symmetry-violating observables scales rapidly with atomic number, so the use of high-Z systems is desirable. Independent, precise atomic wavefunction calculations in these systems are thus necessary in order to distinguish quantum-mechanical effects from the elementary-particle physics observables being targeted. For instance, the present $2\%$-$3\%$ uncertainties in \textit{ab initio} wavefunction calculations in thallium currently limit the quality of the standard model test provided by a 1995 thallium parity nonconservation measurement~\cite{Vetter1995}. Theoretical methods very similar to those used for thallium can be applied to other three-valence systems such as indium and gallium~\cite{Safronova2007}.

Our group recently measured the Stark shift in the 410~nm $5p_{1/2}\rightarrow6s_{1/2}$ transition in indium using an atomic beam~\cite{Ranjit2013}. This measurement was completed with an overall $0.3\%$ accuracy and therefore provided a benchmark test of two distinct calculation methods which can be applied to these multivalence atoms~\cite{Safronova2008}.  Combined with theoretical calculations, this result allowed for precise determinations of the indium $6p_{1/2}$ and $6p_{3/2}$ state lifetimes. Another recent measurement completed by our group targeted the hyperfine constants of the indium $6p_{3/2}$ excited state using a heated vapor cell containing indium~\cite{Gunawardena2009}. This provided a complementary, short-range test of electron wavefunction behavior, and also allowed for determination of the indium nuclear quadrupole moment.

The new polarizability measurement reported here makes use of a two-step excitation in our atomic beam system in order to study the polarizability of the $6p_{1/2}$ excited state in indium. Our result, with its $0.6\%$ uncertainty, is in excellent agreement with recent atomic theory calculations. Together with the theory calculation in Ref.~\cite{Safronova2013} and our previous measurement in Ref.~\cite{Ranjit2013}, the results presented here can be used to deduce a benchmark value for the $6p_{1/2}-5d_{3/2}$ matrix element in indium.

\section{Atomic Structure Details}
\begin{figure}[tb]
\includegraphics[width=1\columnwidth]{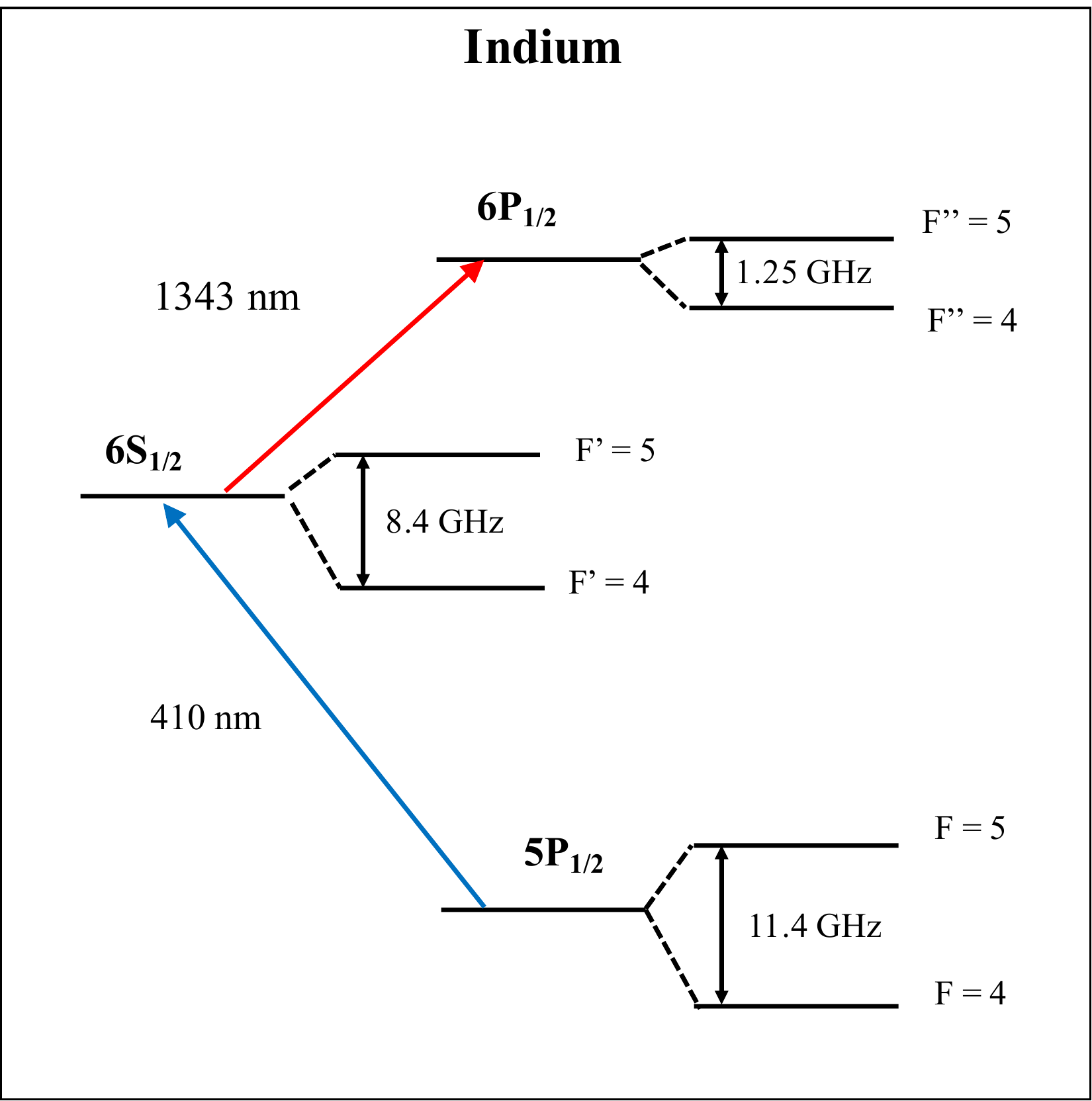}
\caption{Relevant low-lying energy levels of indium. Spacings shown are not to scale.}
\label{fig:IndiumLevels}
\end{figure}

We study the predominant naturally occurring isotope of indium ($^{115}$In, $96\%$ abundance). Our Doppler-narrowed atomic-beam geometry spectroscopically isolates this from the small $^{113}$In component also present. Further details about the relevant atomic structure are discussed in Ref.~\cite{Ranjit2013}. Briefly, $^{115}$In has nuclear spin $I=9/2$, so all three of the $5p_{1/2}$, $6s_{1/2}$, and $6p_{1/2}$ states contain $F=5$ and $F=4$ hyperfine levels (see Figure~\ref{fig:IndiumLevels}). The respective 11.4, 8.4, and 1.2 GHz hyperfine splittings (HFS) of these states yield entirely resolved spectra in our atomic-beam apparatus. Finally, because we only study $J=1/2\rightarrow1/2$ transitions in this work, the Stark shift in each level has only a scalar component; it is therefore independent of the particular hyperfine transition studied, and does not depend on the relative direction between static field and laser beam polarization.  As is discussed in Ref.~\cite{Ranjit2013}, we expect the applied electric field to lead to an energy shift $\Delta E = -\frac{1}{2}\alpha_0 \mathcal{E}^2$, where $\alpha_0$ is the scalar polarizability and $\mathcal{E}$ is the applied electric field. The observed frequency shift in the second-step transition will thus be $\Delta \nu_S = -\frac{1}{2h}\left[\alpha_0(6p_{1/2})-\alpha_0(6s_{1/2})\right]\mathcal{E}^2$ so long as conditions on the detuning of the laser driving the first-step transition outlined below are met. Note that since the difference in polarizabilities $\alpha_0(6s_{1/2})-\alpha_0(5p_{1/2})$ was previously determined by our group to high precision, and $\alpha_0(5p_{1/2})$ is at least an order of magnitude smaller than either of the other polarizabilities, our Stark shift result can be interpreted as a determination of the $6p_{1/2}$-state polarizability with negligible loss of precision.

\subsection{Complications due to two-step excitation}
In a single-step excitation experiment it is clear that the measured Stark shift corresponds to the difference in energy level shifts between the two states being coupled. The problem is more subtle for the case of simultaneously coupled states in a “ladder” configuration. Based on numerical and analytic simulations, we find that the observed shift \textit{is} equal to the difference in Stark shifts of the upper two levels so long as we always tune the first-step laser to satisfy the resonance condition of this transition in the presence of the applied electric field. Our simulations show that if the first-step transition fails to remain at the (Stark-shifted) resonance, the observed field-induced shift in the second-step transition can differ from the expected value by as much as $10\%$ depending on the Rabi frequencies associated with each transition. Therefore, as described below, we have taken the approach of locking the first-step laser to the atomic beam absorption signal itself, guaranteeing that the resonance condition within this transition is satisfied for any value of the static electric field.

\section{Experimental Details}
\subsection{Atomic-beam system}
We make use of the same atomic beam unit described in Ref.~\cite{Ranjit2013}.  A water-cooled in-vacuum furnace heats a molybdenum crucible containing a $\sim$100~g sample of indium to 1200$^{\circ}$C to produce our effusive atomic beam. Even at this temperature, the optical depths of the 410~nm and 1343~nm transitions in our interaction region are of order $10^{-3}$ and $10^{-4}$, respectively. Slits on the output of the crucible, an adjustable set of razor blades, and a rectangular mask near our interaction region serve to collimate the atomic beam and define its profile. The collimation is sufficient to reduce the transverse Doppler broadening in the 410~nm transition by an order of magnitude and ensures that the laser-atom interaction region is restricted to the region in the center of the electric field plates. The collimated atomic beam then passes through a set of parallel, circular capacitor plates producing a static electric field up to 20 kV/cm. Both laser beams intersect the atomic beam in a transverse fashion between the field plates in this interaction region. The atomic beam is housed within a vacuum chamber whose pressure remains at or below $10^{-7}$~torr. Just upstream of the field plates, a chopping wheel modulates the atomic beam at roughly 600 Hz. In order to reduce potential systematic effects due to Zeeman shifts, three pairs of mutually orthogonal magnetic-field-cancellation coils surround the interaction region and reduce ambient magnetic fields here to less than 1~$\mu\text{T}$.

\subsection{Electric field plates and high voltage system}
The electric field plates and high voltage (HV) system used in the present experiment have been described in Ref.~\cite{Ranjit2013}. They consist of two stainless-steel circular plates, each 10 cm in diameter. The two plates are separated by ceramic spacers and leveled horizontally on a ceramic pedestal. With the aid of a CNC milling machine we determine the plate separation to be 0.9999(5) cm. For our plate and beam geometry, non-idealities due to finite field-plate diameter are at the $10^{-5}$ level. We collect data for a range of applied voltages up to 20 kV, monitoring the applied voltage during data acquisition by means of a precisely calibrated voltage divider (Ross Engineering, Inc.) and Keithley (model 197A) voltmeter via GPIB interface.

\subsection{Optical system}
In order to drive the transitions of interest, we make use of two commercial external cavity diode lasers (ECDLs), both operating in the Littrow configuration. One (Toptica Photonics, DL 100) is tuned to either the 410~nm $5p_{1/2}\rightarrow6s_{1/2}(F = 4 \rightarrow F^\prime=5)$ or $(F = 5 \rightarrow F^\prime = 4)$ transition and locked as described below. The other ECDL (Sacher Laser Lynx Series) is scanned across the 1343~nm $6s_{1/2}\rightarrow6p_{1/2}(F^{\prime\prime}=4,5)$ transitions.

\subsubsection{Frequency stabilization of the 410 nm laser}
\begin{figure}[tb]
\includegraphics[width=1\columnwidth]{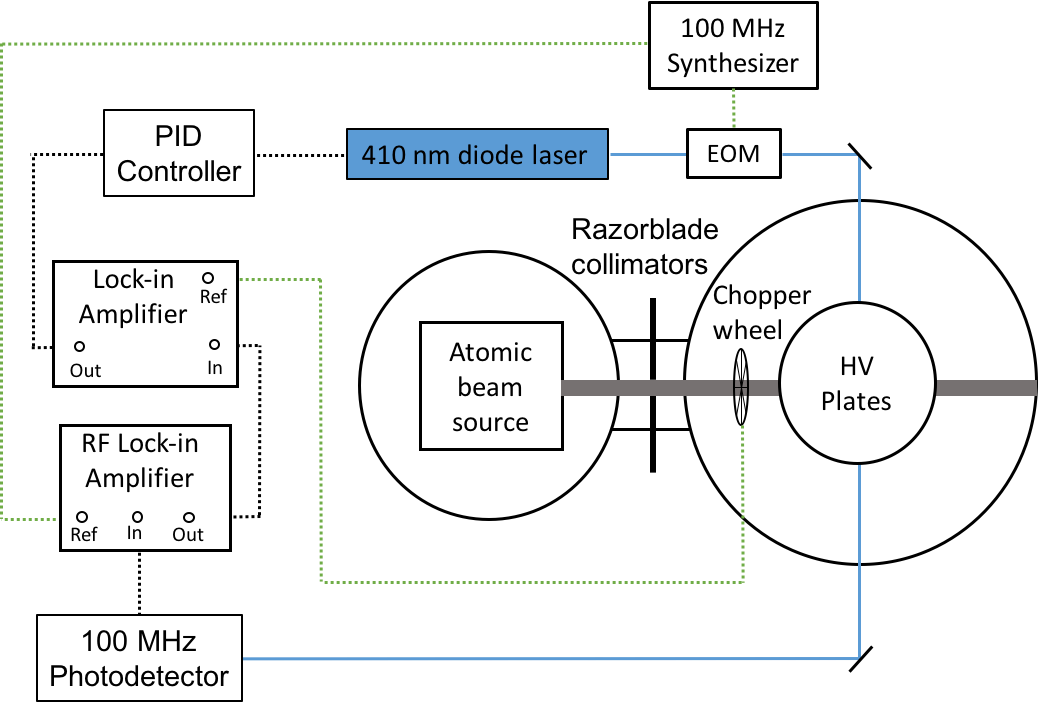}
\caption{A schematic diagram showing the optical and signal processing arrangement for the atomic-beam-based laser locking scheme.}
\label{fig:Locking}
\end{figure}

Immediately after the optical isolator, the blue laser beam passes through a beamsplitter. One component is passed through an electro-optic modulator (ThorLabs EO-PM-NR-C4) driven by a 100 MHz rf synthesizer. As sketched in Fig.~\ref{fig:Locking}, this laser beam is directed into the atomic beam interaction region, intersecting the beam transversely at the midpoint between the electric field plates. This interaction serves as both the basis for our laser locking scheme and as the first of two excitation steps to reach the final $6p_{1/2}$ state. To stabilize the blue laser, we employ frequency-modulation spectroscopy to enhance the signal-to-noise ratio of our very small atomic absorption signal. The transmitted blue light is collected on a fast photodiode (New Focus, model 4001), and we use a rf lock-in amplifier (SRS model SR844) referenced to the EOM drive frequency to produce a characteristic dispersive atomic spectral feature. To further enhance the signal quality and eliminate background drifts not related to the atoms, we pass the output of this lock-in amplifier into a second, low-frequency lock-in amplifier referenced to the atomic beam chopping wheel frequency. This two-step demodulation yields a high signal-to-noise ratio dispersive signal free of DC offset and drift, and suitable for frequency stabilization. A standard \textit{p-i-d} servo circuit locks the laser by feeding back to a  piezoelectric transducer controlling the position of the diffraction grating in the laser's external cavity. We find that over time scales from 10 msec to 1 hour, the residual RMS fluctuations of the locked laser are below 1 MHz. 

To ensure  that the laser lock tracks the changing resonance position as a function of electric field we executed a series of high-voltage steps and monitored the servo circuit correction signal. As expected for a quadratic Stark shift, we observe a four-fold change in the correction signal when we double the high voltage value. For the largest electric fields used here, the Stark shift of the 410~nm transition is roughly 25 MHz, significantly smaller than the Doppler-narrowed $\sim100$ MHz resonance linewidth of the atomic beam signal. We find that the servo system can reliably follow the shifting resonance and re-acquire lock in a fraction of a second. In our data acquisition scheme we typically wait 10 seconds to eliminate transients  after switching the high voltage on or off, so this locking procedure provides a reliable means of tracking and locking to the Stark-shifted resonance.

\subsubsection{Infrared laser and two-tone modulation}
Prior to overlapping the 1343 nm IR laser with the blue laser to perform our two-step excitation we pass the IR optical beam through an EOM resonant at 600~MHz (New Focus, model 4423). We drive the EOM with an rf signal that is the result of mixing the output of a 600 MHz synthesizer with a 50 kHz signal, producing a double set of rf sidebands at 599.95 and 600.05 MHz. Higher order sidebands are of negligible size and do not complicate the spectra. The modulated laser beam then interacts with indium atoms (both in a heated vapor cell and our atomic beam), providing frequency calibration and facilitating rf lock-in detection to enhance the signal-to-noise ratio of the atomic beam signal.

A portion of the IR laser beam is picked off prior to modulation and passes through a confocal Fabry-Perot (FP) cavity (Burleigh RC-110) with free spectral range of 500 MHz and finesse of roughly 20. The cavity is constructed of low-expansion material and is enclosed in an insulated box for passive thermal stabilization. As described below, we use the FP cavity transmission signal to linearize the laser scan prior to analysis of our vapor cell and atomic beam spectra.

\subsubsection{Vapor cell spectroscopy}
In order to provide a field-free reference spectrum for the two-step excitation signal and to monitor this signal in an environment where much higher atomic densities can be produced, we make use of a small table-top furnace which heats a 15-cm-long sealed quartz indium cell to approximately 700$^\circ$C. Prior to the 100 MHz modulation, we direct a portion of our blue laser through the cell and overlap it with a portion of the modulated 1343~nm laser which scans across the indium $6s_{1/2} - 6p_{1/2}$ second-step transition. Using an optical chopping wheel we modulate the blue laser beam incident on the cell at $\sim1$~kHz. We then employ lock-in detection of the infrared transmission signal to obtain background-free IR spectra. Because the blue laser promotes only one velocity class of the Doppler-broadened sample to the intermediate state, our IR spectra are nearly Doppler free. Details of this two-step spectroscopy technique can be found in \cite{Gunawardena2009, Ranjit2014}.

Since the vapor cell remains in an electric field-free environment, the locked blue laser frequency will not remain in resonance with atoms in the vapor cell whenever the atomic beam high voltage is on. To address this, the portion of the 410~nm laser used for vapor cell spectroscopy is first directed through a pair of acousto-optic modulators (AOMs) (Isomet model 1250C-829A, center frequency 260 MHz). These are oriented so as to make use of the $+1$ order beam from the first, and the $-1$ order beam from the second. When driven at the same rf frequency, this arrangement produces no net frequency shift in the output beam. Because we know the value of the Stark shift constant, $k_s$, for this 410~nm transition to high precision\cite{Ranjit2013}, shifting the frequency inputs to the two AOMs by equal and opposite amounts, $ \pm \Delta f = \left(k_s \mathcal{E}^2\right)/2$, exactly compensates for the change in laser frequency caused by application of the electric field $\mathcal{E}$ and ensures that the blue laser beam component in the vapor cell remains centered on the field-free atomic resonance at all times.

\begin{figure}[tb]
\includegraphics[width=1\columnwidth]{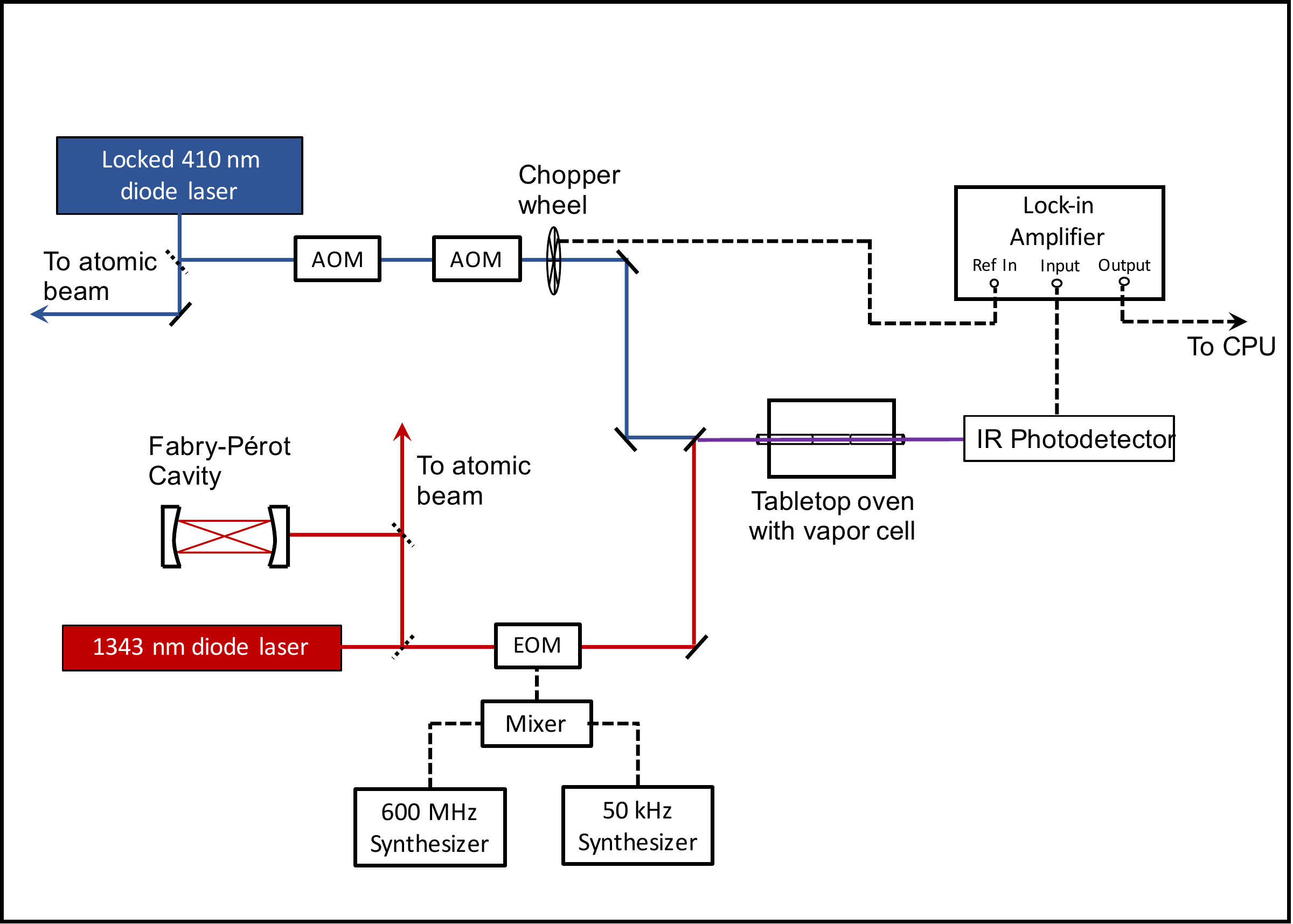}
\caption{A schematic diagram detailing the supplementary indium vapor cell used for obtaining the hyperfine peaks as a reference signal and RF sidebands for frequency calibration.}
\label{fig:VCSetup}
\end{figure}

\begin{figure}[tb]
\includegraphics[width=1\columnwidth]{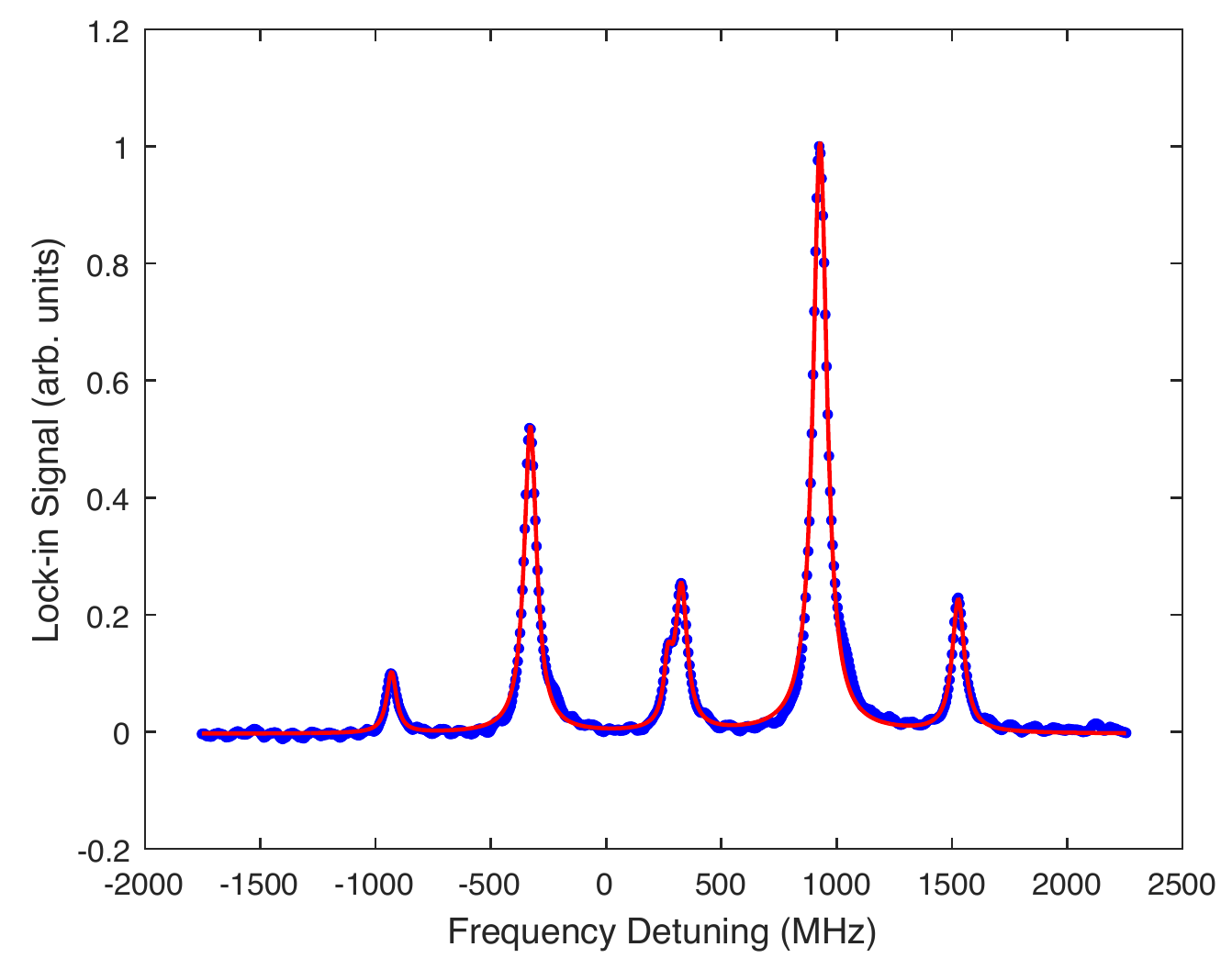}
\caption{A sample HFS reference signal measured via IR light transmitted through a supplementary indium vapor cell. The solid line is a fitted curve using a sum of six Lorentzians. The six peaks include two hyperfine features, each with two sidebands. The central two sidebands are partially overlapped, but can be resolved in the fitting procedure. This signal is used for frequency calibration and referencing as well as to measure the $6p_{1/2}$ hyperfine splitting, as described in the text.}
\label{fig:HFSPeaks}
\end{figure}

Fig.~\ref{fig:VCSetup} shows a schematic diagram of the vapor cell setup. Here we do not employ rf spectroscopy of the IR laser transmission signal, in the sense that we do not demodulate the signal at the rf frequency. Nevertheless, the output of the lock-in amplifier referenced to the blue laser chopping wheel reveals the hyperfine spectrum of the indium $6p_{1/2}$ state along with sidebands at $\pm 600$~MHz relative to each peak. We do not resolve the small $\pm 50$ kHz splitting in our sidebands. A typical spectrum is shown in Fig.~\ref{fig:HFSPeaks}. The sidebands apparent in this figure provide ideal frequency calibration, as described below.

\begin{figure}[tb]
\includegraphics[width=1\columnwidth]{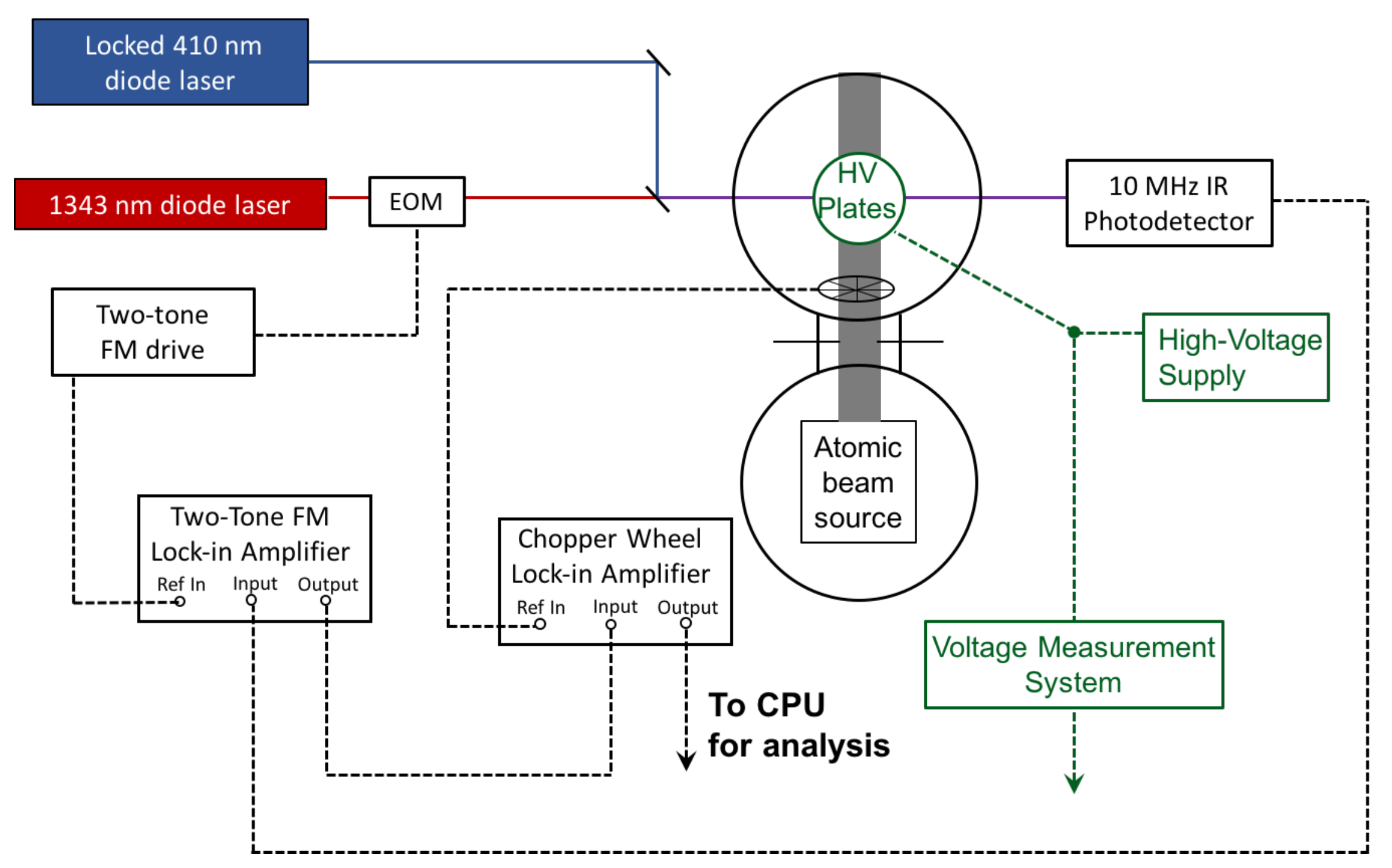}
\caption{A schematic diagram detailing the atomic beam unit setup, including the detection electronics using two successive lock-in amplifiers.}
\label{fig:ABUSetup}
\end{figure}

\subsection{Atomic beam spectroscopy and signal processing}
We direct several mW of blue laser power to the atomic beam for both laser locking and first-step excitation, and overlap this in our interaction region with a comparable amount of IR laser power. The two laser beams are collimated and have a similar beam diameter of roughly 1 mm. Fig.~\ref{fig:ABUSetup} shows a schematic diagram of the atomic beam and interaction region. The stabilized blue laser beam and scanning, frequency-modulated IR laser beam are overlapped on a dichroic mirror and focused to the center of the interaction region. A pair of 1-mm-diameter collimating apertures are placed on either side of the vacuum chamber to ensure a reproducible, well-defined interaction geometry. At the output of the atomic beam unit, the transmitted IR power is detected on a 10-MHz-bandwidth photodiode (New Focus, model 2053).

In order to observe the very small IR absorption in our atomic beam, we implement two-tone frequency modulation (FM) spectroscopy~\cite{Janik1986}. In a two-tone FM scheme, the laser beam is modulated at two nearby frequencies $\omega_m\pm\Omega/2$. It is assumed that $\Omega$ is much smaller than the width of any atomic features of interest, so sidebands separated by $\Omega$ experience the same absorption.  As noted above, $\omega_m = 600$ MHz and $\Omega = $ 100 kHz, satisfying these relative size criteria (residual Doppler broadening in the atomic beam is $\sim100$ MHz). We work at low modulation depth and demodulate the transmission signal at frequency $\Omega$. This yields a spectrum with a central zero-background absorption feature and sidebands which are 180 degrees out of phase and separated by $\pm\omega$~\cite{Janik1986}.

\begin{figure}[tb]
\includegraphics[width=1\columnwidth]{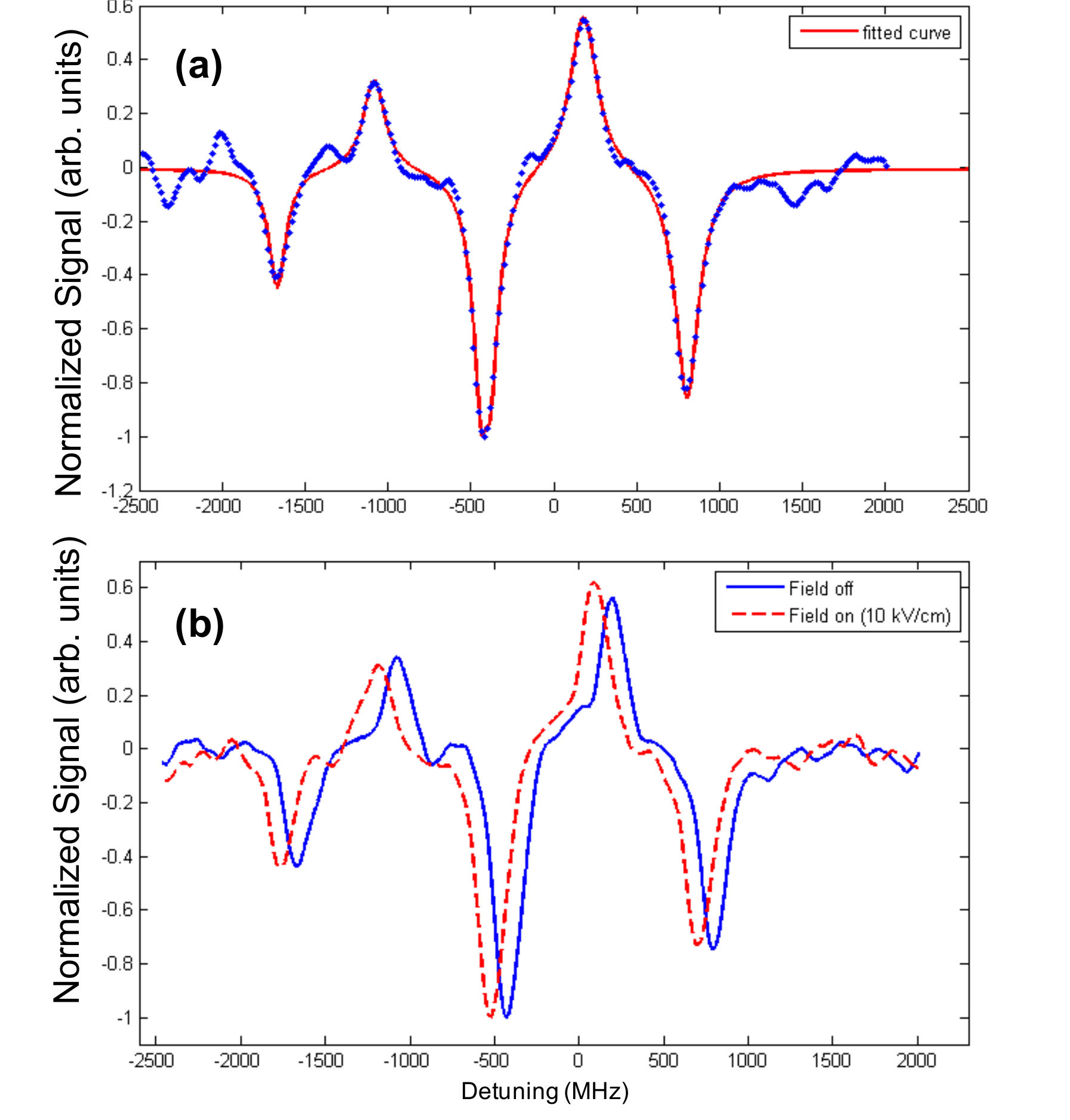}
\caption{(a) Typical two-tone FM demodulated lineshape, obtained by measuring the IR light transmitted through the atomic beam (over a $\sim$4 sec laser scan) and demodulated as described in the text. Fit is to the sum of six Lorentzians.  Negative-going peaks reflect the 180$^\circ$ phase shift of the sidebands in the demodulated spectrum relative to the principal hyperfine peaks. (b) Consecutive field-on and field-off scans, interpolated in order to facilitate the overlap analysis method (see text). In this case the application of a 10 kV/cm electric field results in a roughly 80 MHz Stark shift.}
\label{fig:TTFMSample}
\end{figure}

A particular benefit of two-tone FM spectroscopy is that we may modulate the laser beam at frequencies near 600 MHz, eliminating many sources of noise, while demodulating using standard detection and lock-in electronics at 100 kHz. Although ideally we would expect a  zero-background rf-demodulated signal, optical imperfections in our EOM lead to a small frequency-dependent background pattern independent of the atoms. Hence, we once again make use of the atomic beam chopper wheel and use a second lock-in amplifier referenced to this chopping frequency as a final step in our IR signal detection. This provides a nearly background-free output which we use for data analysis. A typical demodulated IR spectrum is shown in Fig.~\ref{fig:TTFMSample}(a) along with a fit to a sum of Lorenztians.  A pair of consecutive field-off/ field-on scans are shown in Fig.~\ref{fig:TTFMSample}(b), where the $\sim$80 MHz downward Stark shift is easily visible.

\subsection{Data acquisition and experimental control}
Each experimental scan begins by setting the HV state to either on or off. After a HV-state switch, we wait about 10 s for all transients to subside. We then record the IR signal for four linear voltage sweeps of the laser piezo, two upward-going and two downward-going, each of roughly 4 s. Then the HV state is switched once again, and we again record four IR spectra. In this way, we can easily compare Stark shifts obtained from HV on$\rightarrow$off spectra and HV off$\rightarrow$on spectra as a systematic check, and we also require the HV state to switch less frequently. We separately analyze upward-going and downward-going frequency scans as part of our overall effort to explore systematic errors.  As discussed above, each time we switch the state of the HV we apply the appropriate compensating frequency shift to the portion of the 410 nm laser used for frequency referencing frequency. This ensures that the 410 nm laser remains tuned to the field-free first-step resonance in our vapor cell.

\section{Data and Analysis}
\subsection{Summary of data collected}

We collect three experimental signals: the transmitted intensity through the IR FP cavity, the vapor cell HFS reference signal, and the atomic beam demodulated IR signal. A typical experimental run involved recording all of these signals $\sim$20 times at each of about 10 HV values ranging from 7 to 15 kV, resulting in roughly 500 individual spectra and therefore 250 Stark shift determinations per run. For each scan we  record the output of a Keithley voltmeter. Our statistical uncertainty in a single 4 hour dataset such as this is roughly $1\%$.  We collected 10 days worth of such data in all, resulting in roughly 2,500 individual measurements. A range of laser intensities in the interaction region was explored by varying the ratio of blue to IR laser intensity by a factor of 4. We also studied two complementary excitation schemes, using both the $5p_{1/2}(F=5) \rightarrow 6s_{1/2}(F^\prime=4) \rightarrow 6p_{1/2}(F^{\prime\prime}=4,5)$ hyperfine transitions and the $5p_{1/2}(F=4) \rightarrow 6s_{1/2}(F^\prime=5) \rightarrow 6p_{1/2}(F^{\prime\prime}=4,5)$ transitions.

\subsection{Frequency linearization and calibration}
As described in detail in \cite{Ranjit2013, Ranjit2014}, the first step in our data analysis of each particular spectrum involves fitting the FP transmission signal to account for the residual nonlinearities produced in our laser scans.  For our $\sim$5 GHz IR scans, we find such nonlinearities are at about the 1$\%$ level. Fitting the FP spectrum to an Airy function with a frequency parameter parametrized by a fourth-order polynomial adequately accounts for this nonlinearity.

We use the results of these FP fits to generate a linearized frequency axis, and then rely on the sideband splittings in our vapor cell spectral fits to provide absolute frequency calibration for our scans. To achieve this calibration, we fit the vapor cell HFS spectra to a sum of six Lorentzians. It should be noted that, though the Voigt profile is technically the correct line shape to fit our peaks, we see statistically equivalent fit results when using a simpler Lorenztian model, which we have therefore employed throughout. We then re-scale the nominal frequency axis to ensure that the sideband splitting matches the 600 MHz frequency with which we drive the EOM. We find that the two splittings associated with the outer sidebands agree to within 1 MHz.  The small resolved difference places an upper limit on the residual nonlinearity and associated frequency calibration systematic error at the 0.1$\%$ level.  Finally, we reset the zero of the frequency axis to be the position of the $6P_{1/2}(F=5)$ hyperfine peak in the vapor cell signal. Whether the high voltage is on or off in our atomic beam apparatus, the AOM Stark shift compensation scheme maintains a fixed frequency of the blue laser in the vapor cell, allowing the position of the vapor cell IR spectral peaks to serve as a reliable frequency reference point for each atomic beam spectrum.

\subsubsection{$6p_{1/2}$ hyperfine splitting measurement}

With an absolute frequency calibration for our vapor cell IR scans, it is immediately possible to determine the hyperfine splitting between the $6p_{1/2}(F^{\prime \prime}=4,5)$ states in indium. Our several thousand vapor cell scans yield a measurement of this splitting with roughly 0.1 MHz statistical uncertainty. Residual scan nonlinearity in this case limits our final accuracy in HFS determination to 0.5 MHz. We find $\delta\nu_{5-4} = 1257.0(5)$ MHz for this indium $6p_{1/2}$ state; this corresponds to a hyperfine constant of $a_{6p_{1/2}} = 251.4(1)$~MHz. The measured frequency splitting differs by 6~MHz from an older measurement\cite{George1990}, and is a factor of three more precise.

\subsection{Stark shift determination and results}
The final step of the data analysis involves extracting a Stark shift value from each pair of scans. This is done using atomic beam spectra such as those shown in Fig. \ref{fig:TTFMSample}. We have two distinct methods of determining the Stark shift from a pair of HV-on/ HV-off atomic beam spectra. First, we can fit each two-tone FM spectrum to a sum of Lorentzians and measure differences between corresponding peaks in the two spectra. Alternatively, we can use what we term the ``overlap" method, in which we create interpolating functions from each scan and then systematically translate one FM signal across the other, recording a point-by-point sum of squared differences between the signals. The frequency translation at which this difference is minimized must correspond to the point at which the two signals most nearly overlap one another, i.e. where we have compensated for the Stark shift. These two methods are quite complementary in terms of how they make use of the spectral features and have very different potential susceptibility to systematic error.  As will be discussed below, the two Stark shift determination methods agree to better than $0.3\%$. Typically we observe Stark shifts between 20 and 150 MHz, with uncertainties of order 5 MHz in any individual measurement. 

For each set of nominally identical runs at a given electric field we compute a weighted average and standard error of the collected data. We separately analyze data for each laser sweep direction and HV switching direction. By dividing the averaged data by the square of the applied electric field for that particular configuration, we can obtain a results for the Stark shift constant $k_S = \Delta \nu/\mathcal{E}^2$, in units of kHz/(kV/cm)$^2$. This can then be converted into a difference of polarizabilities and, by virtue of the existing measurement and theory for the $5p_{1/2}\rightarrow6S_{1/2}$ Stark shift, into a value for the $6p_{1/2}$ polarizability alone. 

\begin{figure}[tb]
\includegraphics[width=1\columnwidth]{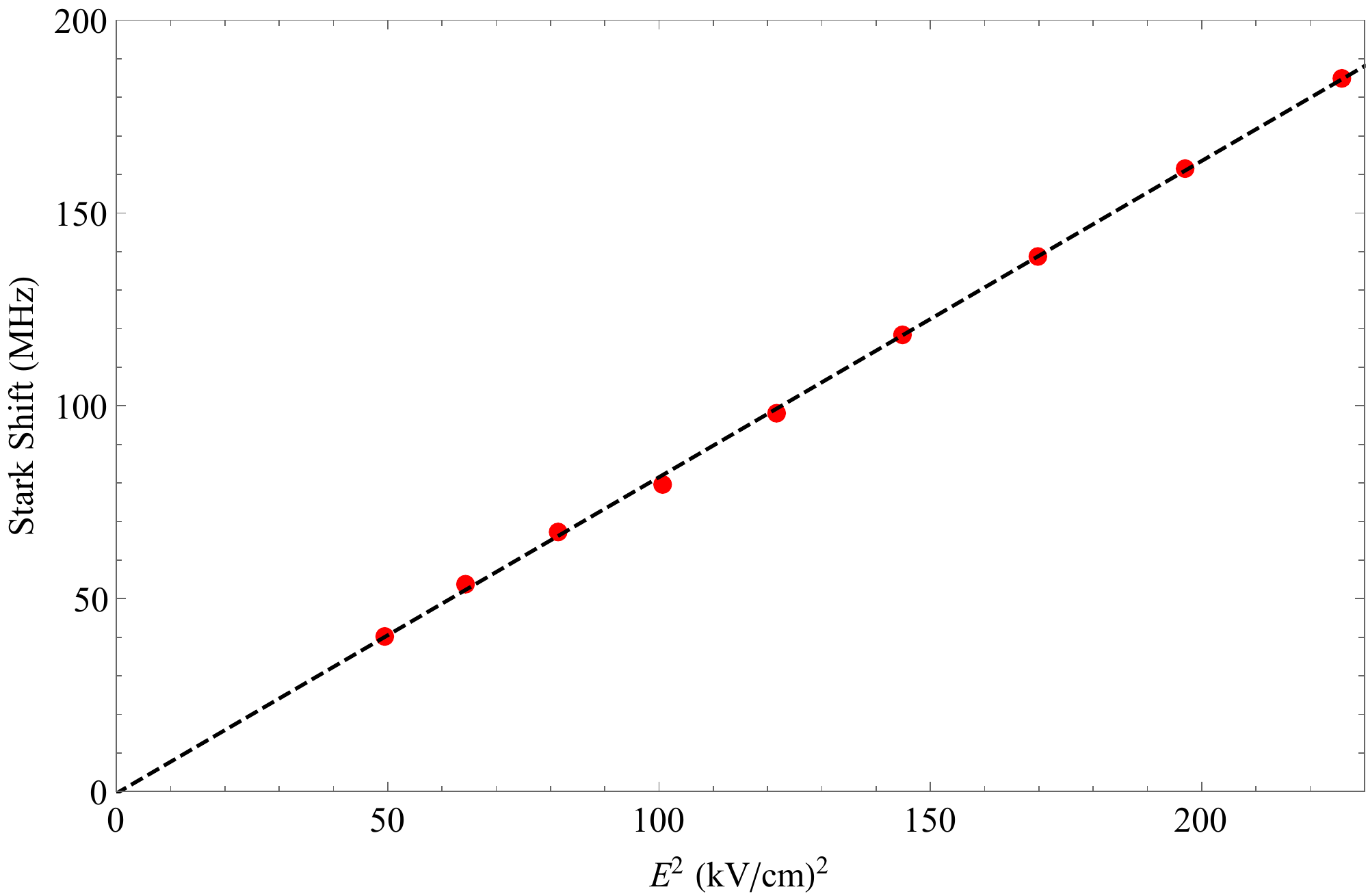}
\caption{Magnitude of measured shift vs. the square of the applied electric field. The result of one day's worth of data is shown, with all runs taken at a given electric field value averaged. Statistical errors of each point are comparable to the point size. The Stark shift constant $ \lvert k_S \rvert $ (slope of fitted line) has an uncertainty of about $1\%$ for these data.}
\label{fig:StarkLine}
\end{figure}

An alternative method of determining $k_S$ is to plot the averaged shift as a function of $\mathcal{E}^2$ and ensure the expected linear dependence, as shown in Fig.~\ref{fig:StarkLine}. As a cross-check to our vapor cell frequency-referencing scheme, we also determine the Stark shifts by using the corresponding FP spectrum as a frequency reference. Given our field on - off- off - on switching sequence, slow thermal drifts in either the cavity length or in the laser external cavity PZT response would be revealed as a systematic difference between consecutive Stark shifts. We observe no resolved drifts of this sort, and find that the two different referencing schemes give results that are in excellent agreement.

\begin{figure}[t]
\includegraphics[width=1\columnwidth]{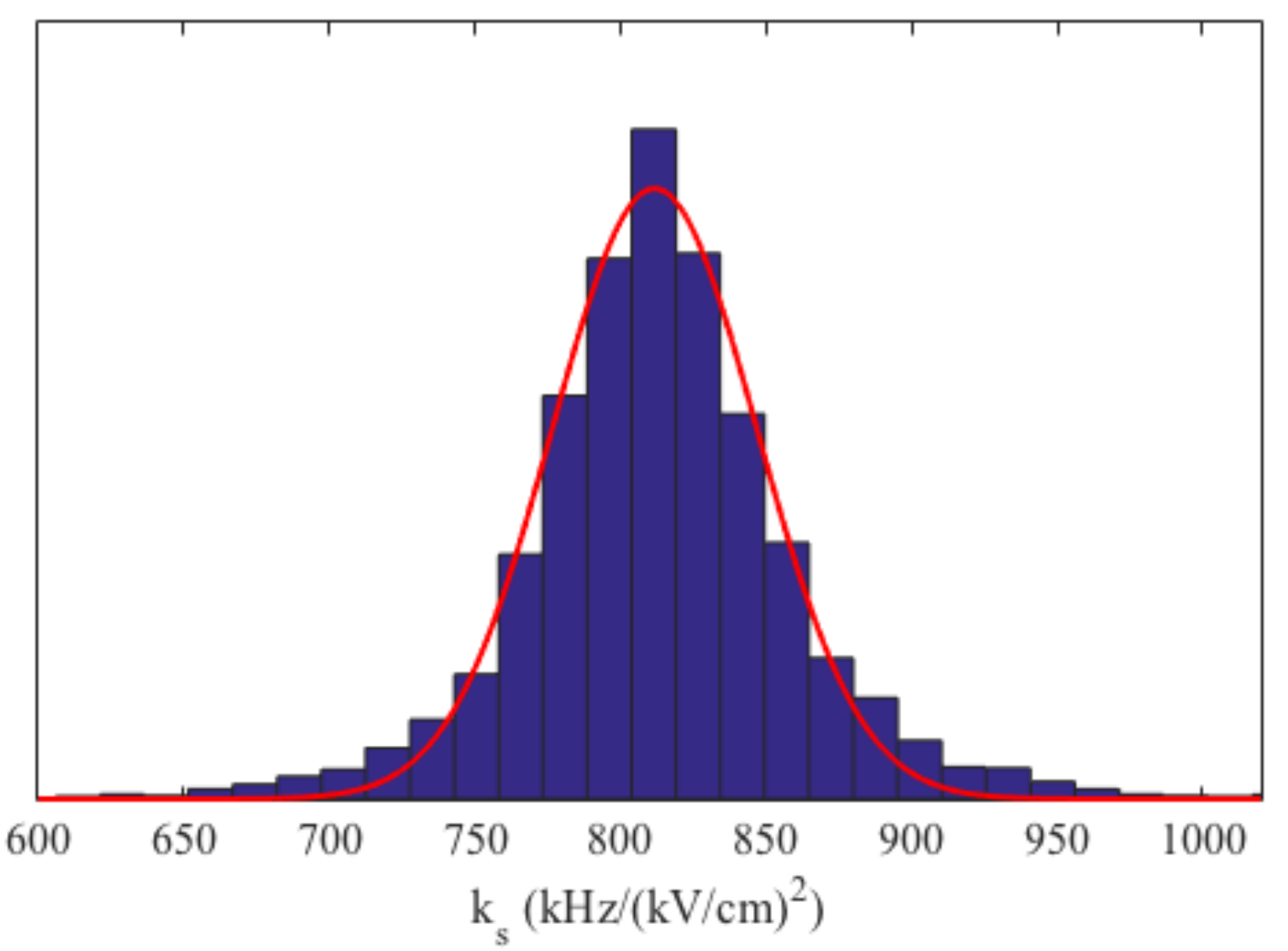}
\caption{A histogram representing all Stark shift measurements taken for electric fields between 7 and 15 kV/cm. The solid line represents a fitted Gaussian.}
\label{fig:PolHist}
\end{figure}

Fig.~\ref{fig:PolHist} shows a histogram of all individual Stark shift constant values obtained from all electric field values.   This histogram represents nearly 3,000 individual Stark shift measurements. We obtained a final value for the Stark shift constant both by studying weighted averages of data at each electric field and by fitting histograms such as shown here. We find good agreement in the values obtained by these methods.  Our final  statistical average is $k_S = - 813.7(2.2)$~kHz/(kV/cm)$^2$, where the error is a one-standard-deviation statistical error.

\subsection{Investigation of systematic errors}
\begin{figure}[t]
\includegraphics[width=1\columnwidth]{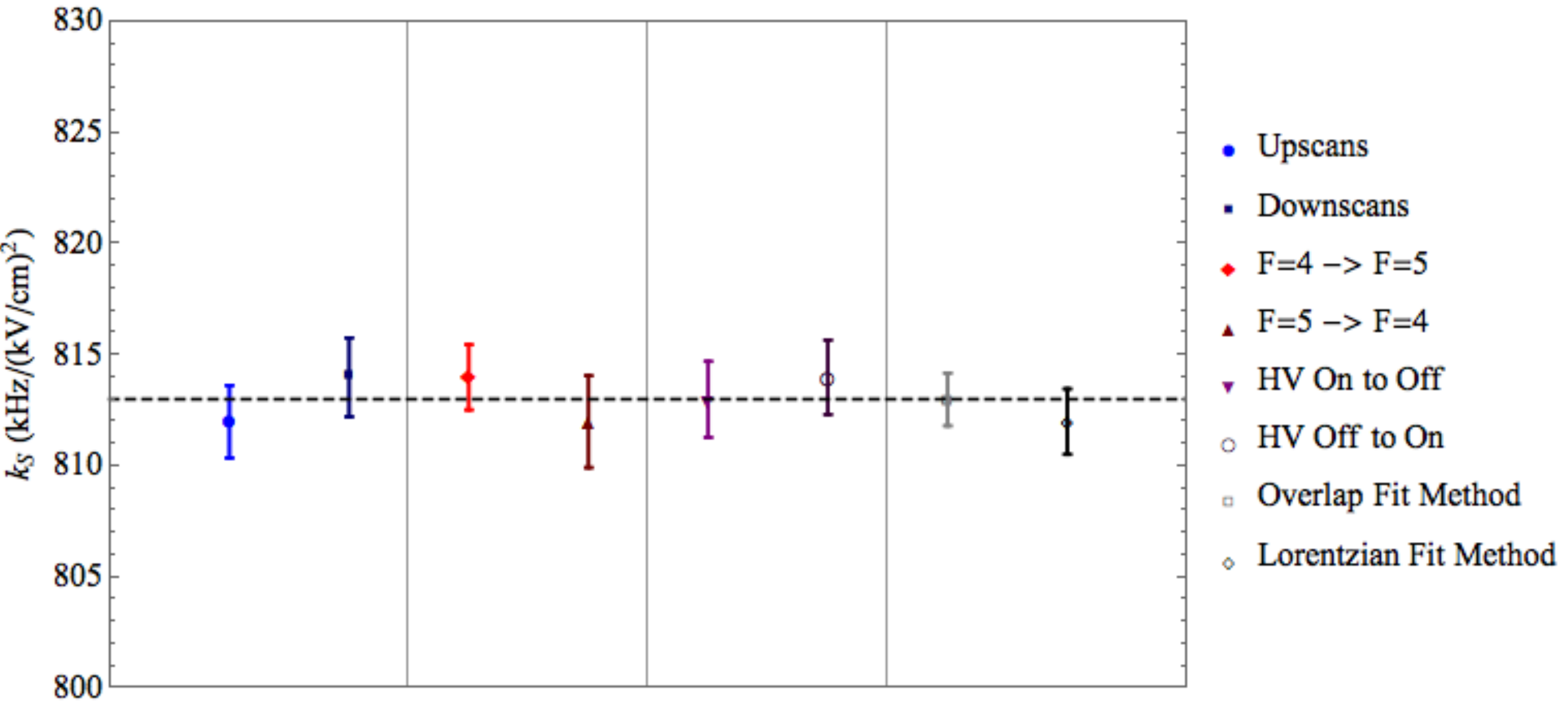}
\caption{Data subset comparisons to investigate potential systematic errors. The comparisons are between: scan direction, intermediate hyperfine transition, HV switching direction, and Stark shift fitting/ extraction method.}
\label{fig:SysError}
\end{figure}

We explored a variety of potential systematic effects, guided by our group's experience with previous Stark shift measurements~\cite{Ranjit2013}. Specifically, we investigated the effect of laser sweep direction (up vs. down), HV switch direction (on$\rightarrow$off vs. off$\rightarrow$on), and Stark shift determination method (peak fitting vs. ``overlap" method). We also compared results obtained for each excited-state hyperfine level. Fig.~\ref{fig:SysError} summarizes the search for systematic errors in such binary comparisons.

\begin{figure}[b]
\includegraphics[width=1\columnwidth]{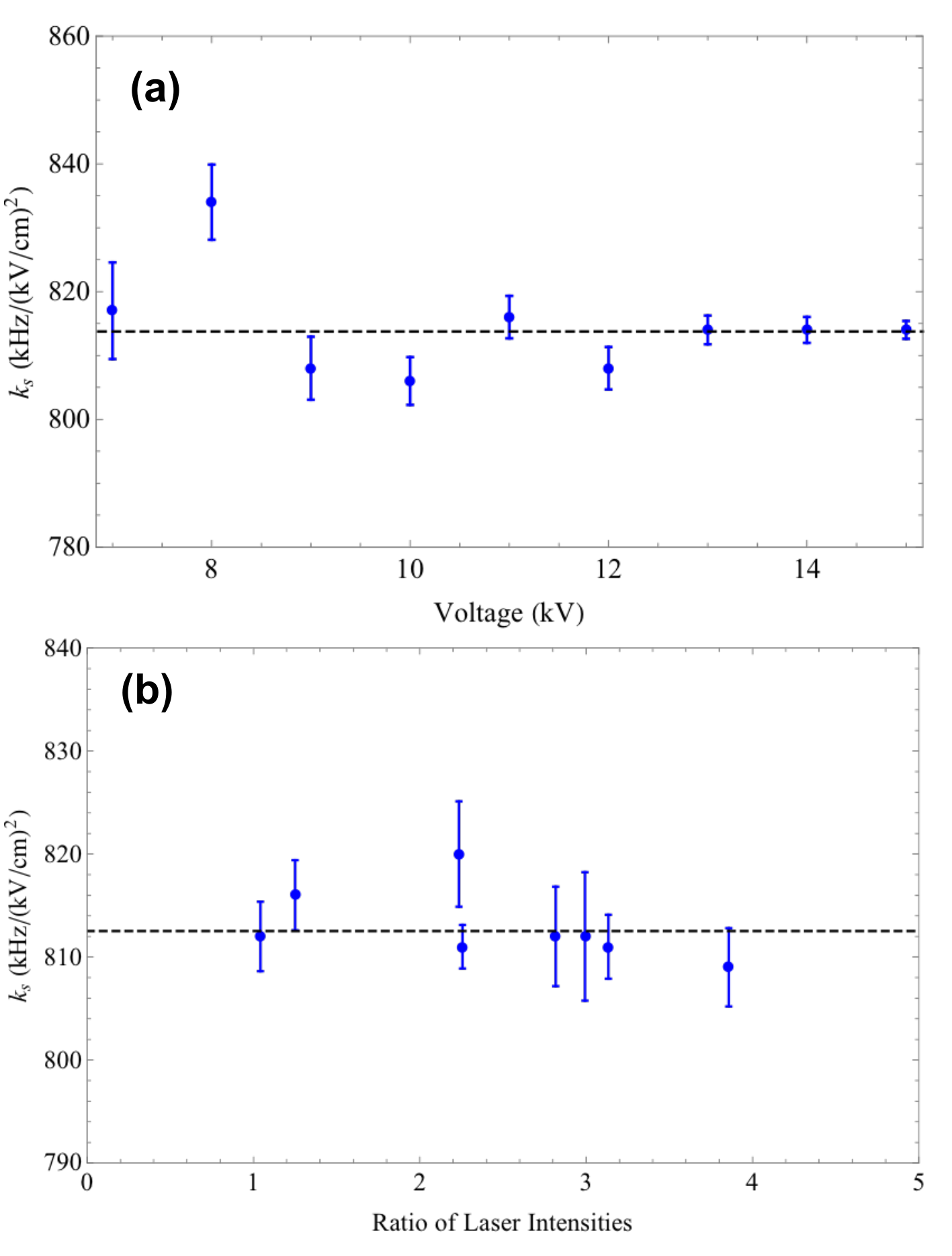}
\caption{(a) Correlation plot of $\vert k_S \rvert $ vs. electric field. No statistically resolved slope is observed. (b) Correlation plot of $\lvert k_S \rvert $ vs. the ratio of blue to IR laser intensities. No statistically resolved slope is observed.}
\label{fig:ShiftVsE}
\end{figure}

As mentioned above, any error in keeping the blue laser locked to resonance would potentially result in a measured Stark shift constant that depended on electric field value and/ or laser intensities due the complications of off-resonant two-step excitation.  Thus, it is particularly important to explore any possible dependence of $k_S$ values on these particular variables. Fig.~\ref{fig:ShiftVsE} shows plots of $k_S$ versus electric field and of $k_S$ versus the ratio of blue to IR laser power, respectively. In neither case do we see statistically significant correlation.  For the case of electric field variation, while we see no statistically resolved slope, we do see statistical fluctuations which exceed those expected given the indicated individual error bars. By considering the $\chi^2$ of the data points in Fig.~\ref{fig:ShiftVsE}(a), we then included a systematic error contribution to account for the observed scatter among the results for various high voltage values. While the aggregated data shown in Figure~\ref{fig:SysError} show no statistically significant systematic differences, a few subsets of data showed marginal statistical disagreement at the level of 1.5 to 2 (combined) standard deviations when compared in this way. In these cases we have included a small systematic error contribution in our final error budget. A full list of systematic error contributions is given in Table~\ref{tab:SysErr}. This table includes a contribution from electric field calibration which aggregates the errors from plate separation measurement and high voltage calibration.  The quadrature sum of these errors, along with the statistical error, yields a 0.6$\%$ final uncertainty in the Stark shift constant.

\begin{table}[tb]
\setlength{\tabcolsep}{12pt}
\begin{centering}
\renewcommand{\arraystretch}{1}
\caption{Summary of results and contributions to the overall error in the measurement of the Stark shift constant for the $6s_{1/2} \rightarrow 6p_{1/2}$ transition. All entries are in units of $\text{kHz/(kV/cm)}^2$.}
\begin{tabular}{lr}
\hline \hline
\textbf{Final result} & $- 813.7$ \\
\hline\hline
\textbf{Statistical error} & 2.2 \\
\hline
\textbf{Systematic error sources:} & \\
Variation with electric field & 2.4 \\
Laser sweep direction & 1.8 \\
Hyperfine transition & 1.4 \\
Fitting/ shift determination method & 1.5 \\
Frequency calibration & 0.9\\
Electric field calibration & 1.3 \\
\hline
\textbf{Combined error total} & 4.5 \\ \hline \hline
\end{tabular}

%

\label{tab:SysErr}
\end{centering}
\end{table}

\section{Discussion of results and comparison with theory}

The value for $k_S$ which we obtain, $k_S = - 813.7(4.5)$ kHz / (kV/cm)$^2$, is easily converted to atomic units. The corresponding atomic parameter is $\alpha_0(6p_{1/2})-\alpha_0(6s_{1/2}) = 6621(43) a_0^3$, the difference in scalar polarizabilities between the upper two states. To determine $\alpha_0(6p_{1/2})$ itself, we first add to this polarizability difference the $6s_{1/2} - 5p_{1/2}$ polarizability difference: $1000 \pm3 \,a_{0}^{3}$ measured recently in our laboratory\cite{Ranjit2013}. Finally, using a recent theoretical calculation~\cite{Safronova2013}, we add the small $5p_{1/2}$-state polarizability of $62 \pm6 \,a_{0}^{3}$ to the total. The small uncertainties in these terms contribute negligibly to the final uncertainty in our experimental determination of $\alpha_0(6p_{1/2})$.   We determine that $\alpha_0(6p_{1/2}) = 7683(43) a_0^3$

Recent theoretical \textit{ab initio} calculations by Safronova \textit{et al.}  of the $6p_{1/2}$ scalar polarizability\cite{Safronova2013}  has produced results in excellent agreement with, though less precise than, our experimental value.  The theoretical uncertainty arises from the difference between two independent methods of polarizability calculation.  The so-called Coupled Cluster (CC) method, which starts with a monovalent approach for the indium atom, yields a value for the $6p_{1/2}$ polarizability of 7817 $a_0^3$.  On the other hand, the Configuration Interaction, all order approach (CI+All) was recently developed by the authors to treat more complicated multivalent systems, such as the heavier group IIIA thallium system\cite{Safronova2013}.  Using this independent approach, the authors obtain 7513 $a_0^3$.  The recommended theoretical value is taken to be the CC result, with an uncertainty of 300 $a_0^3$, reflecting the difference in results of the two methods.  Our result, which sits between the two theory predictions (though favoring the CC value), has the potential to guide the further refinement of these calculations.

 Notably, we can combine our measurement and the theoretical prediction in Ref.~\cite{Safronova2013} to infer a value for the electric dipole (E1) matrix element between the $6p_{1/2}$ state and the nearby $5d_{3/2}$ state. The infinite sum in the theoretical calculation for the $6p_{1/2}$ polarizability is dominated by the term involving mixing between the $6p_{1/2}$ and $5d_{3/2}$ state (nearly 90$\%$ of the total sum). Based on Eq.~3 in Ref.~\cite{Safronova2013}, we can express the scalar polarizability in the valence-electron model as:

\begin{equation}
\alpha_0(6p_{1/2}) = \frac{1}{3} \sum_n \frac{|\langle n||d||6p_{1/2}\rangle |^2}{E_n-E(6p_{1/2})}
\label{eq:SafPol}
\end{equation}

\noindent where $|d|$ is the reduced electric-dipole operator. We can isolate the term contributed by the $5d_{3/2}$ state to this sum, and call the difference between the measured polarizability and the balance of the infinite sum $C$. Then,

\begin{equation}
\langle 5d_{3/2} ||d||6p_{1/2}\rangle = \sqrt{3 |E(6p_{1/2})-E(5d_{3/2})|C}
\label{eq:ImpliedDipole}
\end{equation}

In our case, $\alpha_0(6p_{1/2})=7683(43)\,a_0^3$ and the portion of the infinite sum in the polarizability expression \textit{not} due to the $5d_{3/2}$ state is calculated to be 884(69)$\,a_0^3$~\cite{Safronova2013}. Therefore, we deduce that $C=6799(81)\,a_0^3$. Using the known energy splitting between the $6p_{1/2}$ and $5d_{3/2}$ states, we can deduce a value for the $6p_{1/2} - 5d_{3/2}$ matrix element using Equation~\ref{eq:ImpliedDipole}: $\langle 5d_{3/2} ||d||6P_{1/2}\rangle = 10.00 (6) $ atomic units. This value is in good agreement with recent theoretical calculations, falling between  the CC-based theoretical value of 10.1(1) a.u. and the CI+All method prediction of 9.89 a.u.\cite{Safronova2013}.

\section{Conclusion}
We have measured the scalar polarizability of the $6p_{1/2}$ excited state in atomic indium, and have found good agreement with recent theoretical predictions for this quantity. This measurement was used to deduce the $6p_{1/2}-5d_{3/2}$ matrix element, a result which is also in good agreement with recent calculations. We have recently begun a measurement of the Stark shifts within the $6s_{1/2}\rightarrow7p_{1/2,3/2}$ transitions at 690~nm and 685~nm, respectively. Except for substituting a red diode laser system in place of the IR system, we will make use of the same experimental setup and measurement scheme as described above. Due to the near-degeneracy of the $7p_{1/2}$ and $6d_{3/2}$ states ($\Delta E < 200$~cm$^{-1}$), the polarizability of the $7p_{1/2}$ state is expected to be nearly 40 times larger than the polarizability of the $6p_{1/2}$ state described here. Furthermore, the mixing between $7p$- and $6d$-states should account for about 98\% of this polarizability, allowing very straightforward interpretation of our result in terms of the indium $7p-6d$ matrix element. The $7p_{3/2}$ state will also exhibit a tensor polarizability which provides an additional challenge to atomic theory calculations. 

\begin{acknowledgments}
We thank Nathan Bricault and Nathaniel Vilas for experimental contributions at various stages of this work. Frederick W. Strauch provided important insights into the three-level dynamics described in this paper. We thank Charles Doret for valuable comments on the manuscript. We wish to acknowledge the support of the National Science Foundation RUI program, through Grant No. 1404206.
\end{acknowledgments}


\bibliography{Indium15}

\end{document}